\documentclass[a4paper,14pt]{extarticle}

\usepackage{cmap}
\usepackage[cp1251]{inputenc}
\usepackage[english]{babel}
\usepackage{braket,amssymb,amsmath,array,commath,mathtext,wrapfig,tikz}
\usepackage[arrows=pgf-filled,version=3]{mhchem}
\usepackage{bm}
\usepackage{amsthm,amsfonts,amscd}
\usepackage[super,compress]{cite}
\usepackage{graphicx}
\usepackage{lscape}
\usepackage{geometry}
\usepackage{indentfirst}
\usepackage{afterpage,soul,ulem,dcolumn,changebar,longtable,hhline,multirow,makecell}
\usepackage[small,bf]{caption}
\begin{document}
	
\begin{center}
\textbf{The Negative Energy in Generalized Vaidya Spacetime}

Vitalii Vertogradov

\textit{Herzen State Pedagogical University, St. Petersburg, Russia ; The SAO RAS, Pulkovskoe shosse 65, 196140, Saint-Petersburg, Russia.}

\end{center}
\textbf{Correspondence:} vdvertogradov@gmail.com; Tel.: +7(930)006-15-85

\textbf{Abstract}
In this paper we consider the negative energy problem in generalized Vaidya spacetime. We consider several models when we have the naked singularity as a result of the gravitational collapse. In these models we investigate the geodesics for particles with negative energy when the II type of the matter field satisfies the equation of the state $P=\alpha \rho$ ($\alpha \in [0\,, 1]$).




\section{Introduction}

In 1969 R. Penrose theoretically predicted the effect of the negative energy formation in Kerr metric during the process of the decay or the collision. Further the nature of the geodesics for particles with negative energy was investigated~\cite{bib:grib, bib:ver4}. It was shown that in the ergosphere of a rotating black hole closed orbits for such particles are absent. This geodesics must appear from the region inside the gravitational radius.   Also there was research devoted to the particles with negative energy in Schwarzschild spacetime which was conducted by A.A.Grib and Yu.V.Pavlov~\cite{bib:pavlov}. They showed that the particles with negative energy can exist only in the region inside of the event horizon. However, the Schwarzschild black hole is the eternal one and we must consider the gravitational collapse to speak about the past of the geodesics for particles with negative energy.

Black hole was considered to be the only result of critical gravitational collapse. P. Joshi~\cite{bib:joshi} showed that the result of the gravitational collapse might be the naked singularity (For detailed information see~\cite{bib:rev, bib:ver6}). It means that during the gravitational collapse process the time of the singularity formation is less than the time of the apparent horizon formation and there exists a family of non-spacelike, future-directed geodesics which terminate at the central singularity in the past. M. Mkenyley et al investigated the question about the gravitational collapse of generalized Vaidya spacetime~\cite{bib:mah} and showed that the result of this collapse might be the naked singularity. Further the condition for the mass function were obtained~\cite{bib:ver1, bib:ver2}.  Vaidya spacetime is the one of the earliest examples of cosmic  censorship conjecture violation~\cite{bib:pap}.  Usual Vaidya spacetime has the following form:

\begin{equation}
ds^2=-\left(1-\frac{2M(v)}{r} \right )dv^2+2 \varepsilon dvdr+r^2(d\theta^2+\sin^2\theta d\varphi^2) \,,
\end{equation}
where $v$ represents time and $\varepsilon=\pm 1$ depending on ingoing or outgoing radiation. Vaidya spacetime is known as radiating Schwarzschild spacetime. Here $M$ is the function of time and if the mass doesn't depend on time we have the Schwarzschild spacetime in which the time axis represents the radial null geodesic. If $M$ depends not only on $v$ but also on $r$ then we have generalized Vaidya spacetime.

In usual Vaidya spacetime we have non-zero right hand-side of the Einstein equation. The matter is the  type I of the matter field and represents so-called null dust. The energy momentum tensor has the following form:

\begin{equation}
T_{ik}=\mu L_i L_k \,,
\end{equation}
where $\mu$ is the energy density of the null dust and $L_i$ is a null vector:

\begin{equation}
L_{i}=\delta^0_{i} \,.
\end{equation}
The properties of generalized Vaidya spacetime will be described below.

So if we have the naked singularity formation then we have the question about the negative energy in this case. There is no negative energy in our universe so in the case of naked singularity formation this particles must be forbidden. We consider one explicit model of Vaidya-Anti-de Sitter spacetime when we have the eternal naked singularity and then we consider the negative energy problem in generalized Vaidya spacetime when the  type-II of the matter field satisfies the equation of the state $P=\alpha \rho$ where $\alpha \in [0\,, 1]$.

This paper is organized as follows: In sec. 2 we briefly discuss methods which we use preparing this paper, in sec. 3 we describe the generalized Vaidya spacetime. In sec. 4 we consider the negative energy problem in Vaidya-Anti-de Sitter spacetime. In sec. 5 we consider this problem in the generalized Vaidya spacetime. Sec. 6 is the Discussion.

The system of units $G=c=1$ is used throughout the paper. Also latin indices take values $0\,, 1\,, 2\,, 3$ and greek indices - $1\,, 2\,, 3\,.$

\section{Geodesic Equation, \\ Energy and Angular Momentum.}

In this section we provide necessary method to conduct this research.

This paper is devoted to the geodesics for particles with negative energy. So first of all we should provide the necessary methods to find the geodesic equations and the energy expressions.

The geodesic equation in metric $g_{ik}$ is given by:

\begin{equation} \label{eq:a}
\frac{d^2x^i}{d\lambda ^2}+\Gamma^i_{jk}\frac{dx^j}{d\lambda}\frac{dx^k}{d\lambda}=0 \,,
\end{equation}
where $\Gamma^i_{jk}$ are Cristoffel symbols:

\begin{equation}
\Gamma^i_{jk}=\frac{1}{2}g^{im}\left(g_{jm,k}+g_{mk,j}-g_{jk,m}\right ) \,,
\end{equation}
here sign comma denotes partial derivative:

\begin{equation}
g_{ik,j} =\frac{\partial g_{ik}}{\partial x^j} \,,
\end{equation}
and $g^{ik}$ is contravariant  components of metric tensor:

\begin{equation}
g^{ik}=\frac{g_{IK}}{|g|} \,,
\end{equation}
where $g_{IK}$ is the cofactor of the matrix $g_{ik}$ and $|g|$ is the determinant  of the matrix $g_{ik}$.

The equations \eqref{eq:a} is the second order partial differential equations. But for our purpose it is better to use the first order equations. For this aim we use the lagrangian which is given by:

\begin{equation} \label{eq:laga}
\mathcal{L}=\frac{1}{2}g_{ik}\frac{dx^i}{d\lambda}\frac{dx^k}{d\lambda}  \,.
\end{equation}

We will consider the spherical symmetry spacetime. So in our case we can put $\theta=\frac{\pi}{2}$ and use only three coordinates $\{v\,, r\,, \varphi \}$. Here $v$ is the time. We can obtain the energy and angular momentum expressions using \eqref{eq:laga} i.e.:

\begin{equation} \label{eq:enam}
\begin{split}
-E=\frac{\partial \mathcal{L}}{\partial \dot{v}}=g_{00}\dot{v}+g_{0\alpha}\dot{x}^\alpha \,, \\
L=\frac{\partial \mathcal{L}}{\partial \dot{\varphi}}=g_{22}\dot{\varphi}+g_{2 \alpha}\dot{x}^\alpha \,,
\end{split}
\end{equation}
here dot sign denotes partial derivative with respect to affine parameter $\lambda$.
The expressions \eqref{eq:enam} gives only two equations. To solve this system we need one more equation. We can use the following equation:

\begin{equation} \label{eq:rada}
g_{ik}\frac{dx^i}{d\lambda}\frac{dx^k}{d\lambda}=\delta \,,
\end{equation}
where $\delta =+1\,, -1\,, 0$ denotes timelike, spacelike or null geodesic respectively.

The proof that \eqref{eq:enam} and \eqref{eq:rada}  are the geodesic equations and correspond to \eqref{eq:a} the reader can  see in~\cite{bib:fom}.

We have one important condition - the particles must move in future in time i.e.:

\begin{equation}
\dot{v}=\frac{dv}{d\lambda}>0\,.
\end{equation}

If we satisfy this condition then we can note that if $g_{0\alpha}=0$ in \eqref{eq:enam} then we don't have the negative energy because $-g_{00}$ is always positive. If we consider $g_{00}>0$ then this situation corresponds to the black hole and the particles with negative energy can exist only in the region inside the event horizon.

Thus to have the particles with negative energy outside the event horizon the metric must have off-diagonal term $g_{0 \alpha}\neq 0$. In this paper we consider the generalized Vaidya spacetime which contains off-diagonal term $2dvdr$. So in this case we might expect the particles with negative energy to exist outside the apparent horizon.

As we said above we want to prove the absence of particles with negative energy outside the apparent horizon or in the case of the naked singularity formation in the case of generalized Vaidya spacetime.

We have the naked singularity formation as a result of the continuous gravitational collapse if the time of the apparent horizon formation is more than the time of the singularity formation. Moreover there must exist a family of non-spacelike, future-directed geodesics which terminate at the central singularity in the past.

\section{The Generalized Vaidya Spacetime}

The generalized Vaidya spacetime is widely used to describe the model of gravitational collapse~\cite{bib:mah, bib:mah2, bib:ver1, bib:ver2, bib:ver3, bib:ver5} and the exterior metric of the radiating stars~\cite{bib:l1, bib:l2, bib:l3, bib:l4}.

The generalized Vaidya spacetime has the following form~\cite{bib:vunk}:
\begin{equation}
\begin{split}
ds^2=-\left ( 1-\frac{2M(r,v)}{r} \right ) dv^2+2dvdr+r^2d\Omega^2 \,, \\
d\Omega^2=d\theta^2+\sin^2\theta d\varphi^2 \,,
\end{split}
\end{equation}
here $M(v,r)$  - the mass function depending on coordinates $r$ and $v$ which corresponds to advanced/retarded time.

The generalized Vaidya spacetime differs from usual Vaidya spacetime only by dependence of the mass function on both $r$ and $v$. The right hand-side of the Einstein equations is the mixture of two matter fields type -I and -II. Like in Vaidya spacetime case the type-I of matter field is purely the null dust but the type-II is cosmic strings.

We can write down the energy momentum tensor in the following form:
\begin{equation}
T_{ik}=T^{(n)}_{ik}+T^{(m)}_{ik}\,,
\end{equation}
where the first term corresponds to the matter field type-I and the other one corresponds to the matter field type-II~\cite{bib:hok}.

Now let us write down the expression of the energy momentum tensor:~\cite{bib:vunk}

\begin{equation} \label{eq:ten}
\begin{split}
T^{(n)}_{ik}=\mu L_{i}L_{k}\,, \\
T^{(m)}_{ik}=(\rho+P)(L_{i}N_{k}+L_{k}N_{i})+Pg_{ik} \,, \\
\mu=\frac{2 \dot{M}}{ r^2} \,, \\
\rho=\frac{2M'}{ r^2} \,, \\
P=-\frac{M''}{r} \,, \\
L_{i}=\delta^0_{i} \,, \\
N_{i}=\frac{1}{2} \left (1-\frac{2M}{r} \right )\delta^0_{i}-\delta^1_{i} \,, \\
L_{i}L^{i}=N_{i}N^{i}=0 \,, \\
L_{i}N^{i}=-1 \,.
\end{split}
\end{equation}
here $P$ - pressure, $\rho$ - density  of the II matter field, $\mu$ - the density of the null dust  and $L,N$ - two null vectors.

This model must be physically reasonable so the energy momentum tensor should satisfy weak, strong and dominant energy conditions~\cite{bib:pos}. It means that $\rho$ must be positive and for any non-spacelike vector $v^{i}$:

\begin{equation} \label{eq:emt}
T_{ik} v^{i} v^{k}>0\,,
\end{equation}
and the vector $T_{ik}v^{i}$  must be timelike.
Weak energy conditions also demands that the energy density must be non-negative I.e.

\begin{equation}
\rho >0 \,, \mu > 0 \,.
\end{equation}

The dominant energy condition means that the energy density should not be less than the pressure. If we violate the dominant energy condition then our matter moves along spacelike geodesics which is unphysical.

Strong and weak energy conditions demand:
\begin{equation}
\begin{split} \label{eq:ws}
\mu \geq 0 \,, \\
\rho \geq 0 \,, \\
P\geq 0 \,.
\end{split}
\end{equation}

The dominant energy condition imposes following conditions on the energy momentum tensor:
\begin{equation} \label{eq:dom}
\begin{split}
\mu \geq 0 \,, \\
\rho \geq P\geq 0\,.
\end{split}
\end{equation}

The properties of generalized Vaidya spacetime has been studied for the equation of the state $P=\alpha \rho$ where $\alpha$ belongs to the interval $[0\,, 1]$ in articles~\cite{bib:ver1, bib:ver2}. If we satisfy this equation of the state then the mass function $M(r,v)$ has the form

\begin{equation} \label{eq:mass}
M(r,v)=C(v)+D(v)r^{1-2\alpha}\,,
\end{equation}
where $C(v)$ and $D(v)$ are an arbitrary functions of time $v$. It is also worth noting that the case $\alpha = \frac{1}{2}$ is the special one and we don't consider it here.

When we speak about the naked singularity formation then it means that the time of the singularity formation is less than the time of the apparent horizon formation. And there is a family of non-spacelike future-directed geodesics which terminate at the central singularity in the past.  The singularity forms at the time $v=0$ at $r=0$. So now we should know the equation of the apparent horizon.

To derive the apparent horizon equation we need a tangent vector $u^i$:

\begin{equation}
u^i=-\left ( 1-\frac{2M(v,r)}{r} \right ) \frac{d}{dv}+2\frac{d}{dr} \,.
\end{equation}

Vector $u^i$ tangent to a congruence of outgoing null geodesies. However it doesn't satisfy the geodesic equations

\begin{equation}
u_{i;k}u^k=\xi u_i \neq 0 \,,
\end{equation}
where sign ';' denotes the covariant derivative and

\begin{equation} \label{eq:term}
\xi =2\frac{M(v,r)-rM'(v,r)}{r^2} \,.
\end{equation}

We can't calculate the expansion of the outgoing null geodesies. First of all we need to use an affine parameter $\dot{\lambda}$ and rescaled tangent vector $k^i$:

\begin{equation} \label{eq:vector}
k^i=e^{-\gamma}u^i \,,
\end{equation}
where

\begin{equation}
\frac{\partial \gamma}{\partial \lambda} = \xi (\lambda) \,,
\end{equation}
 where $\lambda$ is an original affine parameter. To calculate the expansion $\Theta$ which is

\begin{equation} \label{eq:ex}
\Theta = k^i_{;i} \,,
\end{equation}
now using \eqref{eq:term}, \eqref{eq:vector} and \eqref{eq:ex} we can write down the expansion expression:

\begin{equation}
\Theta =2\frac{e^{-\gamma}}{r^2} \left (1-\frac{2M(v,r)}{r} \right ) \,.
\end{equation}
Note that the term $2\frac{e^{-\gamma}}{r^2}$ doesn't have impact on the sign of $\Theta$. Finally we have the apparent horizon equation

\begin{equation}
1-\frac{2M(v,r)}{r}=0 \,.
\end{equation}

\section{The Negative Energy in Vaidya-Anti-de Sitter spacetime}

First of all, we decided to consider the energy problem in Vaidya-Anti-de Sitter spacetime because in this case we have:

\begin{itemize}
\item The eternal naked singularity i.e. the singularity will be never covered with the apparent horizon,
\item Under some conditions we have two apparent horizons,
\item at the limit $\lim\limits_{r \to \infty}$ our spacetime becomes Minkowski spacetime and like in Kerr black hole the energy which we consider is the energy  with regard to infinity.
\end{itemize}

In the case of Vaidya-Anti-de Sitter spacetime the type-II of the matter field satisfies the equation of the state:

\begin{equation}
P=\rho \,.
\end{equation}
it means $\alpha=1$ and using \eqref{eq:mass} we obtain:

\begin{equation}
M(v,r)=C(v)+D(v)r^{-1}  \,.
\end{equation}
Here we must consider the function $D(v)$ properly. It can't be positive because the density expression is given by:

\begin{equation}
\rho= -\frac{D(v)}{r^4} < 0 \,,
\end{equation}
and if $D(v)>0$ then we violate energy conditions \eqref{eq:ws} and \eqref{eq:dom}. So to satisfy these conditions we must assume that $D(v)<0$ or we can write:

\begin{equation}
D(v)=-D'(v) \,,
\end{equation}
where now function $D'(v)$ is positive.

We want to investigate the question about the singularities in this spacetime which are formed at $v=0$ and $r=0$. But in this case we must impose some  extra conditions on functions $C(V)$ and $D'(v)$ not to violate energy conditions \eqref{eq:ws} and \eqref{eq:dom} because if $\dot{C}(v)<\frac{\dot{D'}(v)}{r}$ at some point then $\mu$ becomes negative and we violate our energy conditions. So we must consider two cases:

\begin{enumerate}
\item $D'(v)\equiv \mu$, where $\mu$ is positive real constant.  In this case $\dot{D'}(v)\equiv 0$ and if we impose the following condition $C(V) \geq 0$ then we satisfy  all energy condition.
\item $D'(0)=0$ but in this case even if $\lim\limits_{v\to 0, r\to 0} \frac{D'(v)}{r}=X_1$ where $X_1$  finite positive constant and if $C(v)>X_1$ then anyway we would violate energy conditions at the later stage.
\end{enumerate}

So the Vaidya-Anti-de Sitter spacetime is given by:

\begin{equation}
ds^2=-\left( 1-\frac{2C(v)-2\mu r^{-1}}{r}\right )+2dv dr+r^2(d\theta+\sin^2\theta d\varphi^2 ) \,.
\end{equation}
The apparent horizon equation in the case of Vaidya-anti-de Sitter spacetime is given by:

\begin{equation}
\frac{2C(v)}{r}-\frac{2\mu}{r^2}-1=0 \,.
\end{equation}
The time of the singularity formation is $v=0$. So solving the above equation we obtain:

\begin{equation} \label{eq:vad}
\begin{split}
r^2-2C(v)r+2\mu =0 \,, \\
\frac{D}{4}=C^2(v)-2\mu \,.
\end{split}
\end{equation}
From this equation we can conclude that if

\begin{equation} \label{eq:conenf}
C^2(v)< 2\mu \,,
\end{equation}
then the apparent horizon will be never formed.

In this paper we will not prove that the radial null geodesics can terminate at the central singularity in the past.

To obtain the energy expression we use the lagrangian:

\begin{equation}
\mathcal{L}=\frac{1}{2} g{ik}\dot{x}^i\dot{x}^k \,.
\end{equation}
In Vaidya-Anti-de Sitter spacetime the Lagrangian is given by:

\begin{equation} \label{eq:lagva}
2\mathcal{L}=-\left(1-\frac{2M(v,r)}{r}\right )\dot{v}+2\dot{v}\dot{r} +r^2(\dot{\theta}+\sin^2\theta \dot{\varphi}^2) \,.
\end{equation}
Due to spherical symmetry we can put $\theta=\frac{\pi}{2}$ and $\dot{\theta}=0$.

Using \eqref{eq:lagva} we can obtain the energy and angular momentum expressions:

\begin{equation} \label{eq:en}
\begin{split}
-E(v)=\frac{\partial \mathcal{L}}{\partial \dot{v}} =-\left (1-\frac{2C(v)-2\mu r^{-1}}{r}\right ) \dot{v} +\dot{r} \,, \\
L=\frac{\partial \mathcal{L}}{\partial \dot{\varphi}}=r^2\dot{\varphi} \,,
\end{split}
\end{equation}
here $E(v)\,, L$ are energy and angular momentum respectively and dot represents the partial derivative with respect to affine parameter $\lambda$.

The radial geodesic is given by

\begin{equation} \label{eq:geodesic}
\dot{r}=\pm\sqrt{E^2(v)-\left(1-\frac{2C(v)-2\mu r^{-1}}{r}\right ) \left(\frac{L^2}{r^2}-\delta \right )}\,,
\end{equation}
 where $\delta=-1$  in the case of timelike geodesics and $0$ in the case of null ones.

The Vaidya-Anti-de Sitter spacetime contains the eternal naked singularity if we can satisfy the condition \eqref{eq:conenf} in this case $\Theta$ is always positive. We also have the condition $\dot{v}>0$ because we must go to future in time. So the first term in the energy expression

\begin{equation}
E(v)=\left( 1- \frac{2C(v)-2\mu r^{-1}}{r} \right ) \dot{v}-\dot{r} \,,
\end{equation}
is always positive. Thus to obtain the negative energy $\dot{r}$ must be positive. If the negative energy exist it must move from the center to infinity. Due to these conditions and the equation \eqref{eq:geodesic} we obtain:

\begin{equation} \label{eq:cond}
\left (1-\frac{2C(v)-2\mu r^{-1}}{r}\right ) \dot{v} =E-\sqrt{E^2(v)-\left(1-\frac{2C(v)-2\mu r^{-1}}{r}\right ) \left(\frac{L^2}{r^2}-\delta \right )}\,.
\end{equation}

Let's look at the expression which is under the root. We can rewrite it in the form

\begin{equation} \label{eq:root}
E^2+g_{00}\left(\frac{L^2}{r^2}-\delta \right ) \,.
\end{equation}

 The expression in round bracket is positive i.e.
$\delta=0$ or $\delta=-1$, $g_{00}<0$ is the condition for the positivity of the expansion $\Theta$. So $g_{00}\left( \frac{L^2}{r^2}-\delta \right ) <0$. Due to this, the whole expression \eqref{eq:root} is less than $E^2$. It means that the right hand-side of the equation \eqref{eq:cond}

\begin{equation}
E^2-|\sqrt{E^2+g_{00}\left( \frac{L^2}{r^2}-\delta \right)}|_{<|E|}<0 \,,
\end{equation}
is negative. So the negative energy in the case of the naked singularity formation is possible only if we violate the condition $\dot{v}>0$.
So we can conclude that there is no non-spacelike geodesics for particles with negative energy in the case of the naked singularity. In the next part we will show that in generalized Vaidya spacetime if we have the naked singularity formation that the particles with negative energy are forbidden.

Now let's consider the violation of the condition \eqref{eq:conenf}. In this case the equation \eqref{eq:vad} has two different roots:

\begin{equation}
r_{\pm }=C(v)\pm \sqrt{C^2(v)-2\mu} \,.
\end{equation}

In this case we have two different apparent horizons at $r=r_-$ and $r=r_+$.  The expansion $\Theta$ is positive in two regions:

\begin{equation}
\begin{split}
0 \leq r < r_- | \Theta >0 \,, \\
r_-<r<r_+ | \Theta <0 \,, \\
r_+<r<\infty | \Theta>0 \,.
\end{split}
\end{equation}

Again if we consider the particles with negative energy in the regions where $\Theta$ is positive then we are in the previous case and such particles are forbidden. If we consider the region where $\Theta$ is negative then we should put $g_{00}>0$. So let's consider this case. The energy expression has the same form as \eqref{eq:en}:

\begin{equation}
E(v)=\left( 1-\frac{2C(v)-2\mu r^{-1}}{r}\right ) \dot{v}-\dot{r} \,.
\end{equation}

In this case we have two possibilities:

\begin{enumerate}
\item $\dot{r}>0$ the movement from the singularity to infinity,
\item $\dot{r}<0$ the movement to the singularity.
\end{enumerate}

Let's consider the first case when $\dot{r}>0$. As we showed in previous case the expression \eqref{eq:root} is less than $E^2$ but in this case $g_{00}>0$ and it means that the expression \eqref{eq:root} is bigger than $E^2$. Hence the right hand-side of \eqref{eq:cond} is more than zero. And it means that again the non-spacelike geodesics for particles with negative energy are forbidden due to the violation of the condition $\dot{v}>0$.

Now let's return to the second case $\dot{r}<0$. Here we have:

\begin{equation}
\left ( 1-\frac{2C(v)-2\mu r^{-1}}{r}\right ) \dot{v}=E(v)-\sqrt{E^2+g_{00}\left (\frac{L^2}{r^2}-\delta \right )} \,.
\end{equation}
We can see that the right hand-side of the expression above is always negative. In this case the non-spacelike geodesics for particles with negative energy are not forbidden.

Now let's consider the question about the existence of circular or closed non-circular orbits. Both cases demand that the effective potential must be zero at some point $r_0$. First of all let's define the effective potential which is given by:

\begin{equation} \label{eq:sb}
v_{eff}=-(\dot{r})^2=-\left [E^2+g_{00}\left (\frac{L^2}{r^2}-\delta \right ) \right] \,.
\end{equation}

We know that $E<0$ and both $g_{00}$ and $\frac{L^2}{r^2}-\delta$ are positive. So the whole expression in square brackets \eqref{eq:sb} are positive and strictly more than zero. Thus there is no any $r_0$ that $v_{eff}(r_0)=0$.  So in this case the closed non-circular and circular orbits are absent.

We showed that in the region $r_-<r<r_+$ the particles with negative energy can exist. However in this particular case this is unphysical because we demand $g_{00}>0$ and it means that all components of the metric tensor i.e. $g_{00}\,, g_{01}\,, g_{22}\,, g_{33}$ are positive and the line element is spacelike in this region.  To have the non-spacelike line element we must change the sign in front of the off-diagonal term and consider not falling matter but radiation. We consider this case in the section below.

\section{The Negative Energy in Generalized Vaidya Spacetime}

In the case of generalized Vaidya spacetime we will consider only the case when the  type-II of the matter field satisfies the equation of the state $P=\alpha \rho$ where $\alpha \in [0\,, 1]$. In this case we can satisfy all energy conditions \eqref{eq:ws}, \eqref{eq:dom}.

The result which was obtained in the previous section is valid in generalized Vaidya spacetime e.g. if we have the naked singularity formation then the negative energy is impossible. Let's prove it.

The line element we write down in the following form:

\begin{equation}
ds^2=-\left(1-\frac{2M(v,r)}{r}\right ) dv^2+2\varepsilon dv dr+r^2 (d\theta^2+\sin^2\theta d\varphi^2) \,,
\end{equation}
here $\varepsilon=\pm 1$ corresponding ingoing or outgoing matter respectively.

The energy and the angular momentum has the following form:

\begin{equation} \label{eq:engv}
\begin{split}
E(v)=\left ( 1 - \frac{2M(v,r)}{r} \right ) \dot{v}-\varepsilon \dot{r} \,, \\
L=r^2\dot{\varphi} \,.
\end{split}
\end{equation}
The radial geodesic is given by:

\begin{equation} \label{eq:radgv}
\dot{r}^2=E(v)^2-\left ( 1 - \frac{2M(v,r)}{r}\right ) \left( \frac{L^2}{r^2} -\delta \right ) \,.
\end{equation}
Note that the radial equation doesn't depend on $\varepsilon$.

Now we consider the case of the naked singularity formation. It means that $g_{00}<0$. Let's start with the case $\varepsilon=+1$. In this case the first expression \eqref{eq:engv} becomes:

\begin{equation}
E(v)=\left(1-\frac{2M(v,r)}{r}\right )\dot{v}-\dot{r} \,.
\end{equation}
Here the negative energy can exist only in the case of $\dot{r}>0$. It gives:

\begin{equation} \label{eq:coin}
\left(1-\frac{2M(v,r)}{r}\right )\dot{v}=E(v)+\sqrt{E^{2}(v)+g_{00}\left(\frac{L^2}{r^2}-\delta\right)} \,.
\end{equation}

Like in previous section the expression under the root is less than $E^2(v)$ (due to the fact that $g_{00}<0$). Hence it means that the right hand-side of the equation above is less than zero. However the left hand-side must be always positive ($-g_{00}\dot{v}$) and it can be negative only in the case of violating the main principle $\dot{v}>0$. So in the case $\varepsilon=+1$ and the naked singularity formation the particles with negative energy are forbidden.

Now let's consider the second case $\varepsilon=-1$. In this case we have:

\begin{equation}
E=-g_{00}\dot{v}+\dot{r}\,.
\end{equation}
In this case the only possibility for the negative energy to exist is the condition $\dot{r}<0$. It gives:

\begin{equation}
-g_{00}\dot{v}=E+\sqrt{E^2+g_{00}\left (\frac{L^2}{r^2}-\delta \right )} \,.
\end{equation}
This equation coincides with \eqref{eq:coin} above. It means that in general case if we have the naked singularity formation the particles with negative energy are forbidden.

Now let's consider the case $g_{00}>0$ i.e. the black hole formation. Note, that in this case $\varepsilon$ must be equal to $-1$ otherwise the line element will be spacelike which is unphysical.

The first expression of \eqref{eq:engv} becomes:

\begin{equation}
E=-g_{00}\dot{v}+\dot{r} \,.
\end{equation}
Here we have two possibilities $\dot{r}>0$ and $\dot{r}<0$.
Let's consider the first case. When $\dot{r}>0$ we have the negative energy if:

\begin{equation}
-g_{00}\dot{v}=E(v)-\sqrt{E^2(v)+g_{00}\left (\frac{L^2}{r}-\delta \right )} \,.
\end{equation}

Now let's prove that there is no circular and elliptical orbits inside the apparent horizon. Under the notion 'elliptical' we mean non-circular closed orbits. The circular orbit can exist if:

\begin{equation}
\begin{split}
v_{eff}(r_0)=0 \,, \\
\frac{d v_{eff}(r)}{dr}|_{r=r_0}=0 \,,
\end{split}
\end{equation}
here $v_{eff}$ is the effective potential. In our case the effective potential has the form

\begin{equation}
v_{eff}=-\dot{r}^2=-\left[E^2+g_{00}\left ( \frac{L^2}{r^2}-\delta \right ) \right ] \,.
\end{equation}

From \eqref{eq:radgv} we can see that $E^2>0\,, -g_{00}<0$ and $\frac{L^2}{r^2}-\delta >0$ and because $E(v)\neq 0$ the condition $v_{eff}(r_0)=0$ is impossible. The elliptical orbits also has the condition $v_{eff}(r_0)=0$. So in this case we can conclude that there is no circular and elliptical orbits for particles with negative energy.

Now let's consider the second case when $\dot{r}<0$. In this case we have:

\begin{equation}
-g_{00}\dot{v}=E+\sqrt{E^2+g_{00}\left ( \frac{L^2}{r^2}-\delta \right ) } \,.
\end{equation}
Again the expression under the root is bigger than $E^2$ due to $E <0 \,, E\neq 0$ and positivity of $g_{00}$ and $\frac{L^2}{r^2}-\delta$. And it means we can't satisfy the condition $\dot{v}>0$. So in this case the particles with negative energy are forbidden.

\section{Discussion}

The main goal of this article is to prove the absence of the particles with negative energy in the case of the naked singularity formation in the generalized Vaidya spacetime. We have considered one explicit example of Vaidya-Anti-de Sitter spacetime and proved that in the case of naked singularity formation the particles with negative energy are forbidden. Then we have considered the general case and also proved the absence of the particles with negative energy in the case of the naked singularity formation. 

According to the Penrose process the particles with negative energy can exist in the ergoregion of a rotating black hole. Geodesics for such particles have been considered in~\cite{bib:grib}. It has been shown that this type of geodesic can't cross the static limit and these particles are forbidden in the region outside of the static limit. These geodesics have been also studied in the case of Schwarzschild spacetime~\cite{bib:pavlov}. In this metric the particles with negative energy can exist only in the region inside the event horizon and are forbidden outside it. But in the case of the naked singularity formation there is no the apparent horizon and such particles can exist in our universe. Vaidya spacetime is one of the first examples of the naked singularity formation. We hasn't considered the case of usual Vaidya spacetime in this paper because it a particular case of the generalized Vaidya spacetime and the results were obtained in this article are also valid in the case of usual Vaidya metric.

Due to off-diagonal term in generalized Vaidya spacetime there is a possibility for the existence of the particles with negative energy but they can exist only in the region inside the apparent horizon. If the horizon is absent then we can have such particles only if they move back in time which is unphysical. 

There are many models of the gravitational collapse when the result is the naked singularity (See for example~\cite{bib:joshi}). The physics demands that the particles with negative energy can not exist in the case of the naked singularity formation. If they existed then the distant observer would be able to detect but it is forbidden. In these models if the metric doesn't have off-diagonal term then due to the fact that $g_{00}<0$ the particles with negative energy can not exist. If the metric has off-diagonal term $g_{0\alpha} \neq 0$ then these metrics should be investigated to prove the absence of such particles in the case of the naked singularity formation. The generalized Vaidya spacetime is one of such type of metrics and as we have proven the negative energy for this metric is forbidden in the case of the naked singularity formation.

\textbf{acknowledgments} The author says thanks to professor Pankaj Joshi for scientific discussion. This work was supported by RFBR grant 15-02-06818-a.The work was performed within the SAO RAS state assignment in the part "Conducting Fundamental Science Research".

\end{document}